\documentclass[twocolumn,aps]{revtex4}
\usepackage{bm}
\begin{document}

\newcommand{\ket}[1]{\left|#1\right\rangle}
\newcommand{\bra}[1]{\left\langle#1\right|}

\title{Low-temperature Dephasing and Renormalization in Model Systems}

\author{Dmitri S. Golubev$^{1,3}$, Gerd Sch\"on$^{1,2}$, and Andrei
D. Zaikin$^{2,3}$} 
\affiliation{$^1$Institut f\"ur Theoretische Festk\"orperphysik,
Universit\"at Karlsruhe, 76128 Karlsruhe, Germany \\
$^2$Forschungszentrum Karlsruhe, Institut f\"ur Nanotechnologie,
76021 Karlsruhe, Germany\\
$^3$I.E.Tamm Department of Theoretical Physics, P.N.Lebedev
Physics Institute, 119991 Moscow, Russia}

\begin{abstract}

We investigate low-temperature dephasing in several model systems,
where a quantum degree of freedom is coupled to a bath.  
Dephasing, defined as the decay of the
coherence of inital non-equilibrium states,
also influences the dynamics of equilibrium correlation
and response functions, as well as static interference effects. In 
particular in the latter case dephasing should be
distinguished from renormalization effects. 
For illustration, and because of its relevance for
quantum state engineering in dissipative environments,
we first reconsider dephasing in spin-boson models.
Next we review Caldeira-Leggett models, with applications, e.g., to
persistent currents in mesoscopic rings. 
Then, we analyze the more general problem of a particle  which
interacts with a quantum field 
$V(t,{\bf r}(t))$, the fluctuations of which are characterized
by a dielectric function $\epsilon(\omega,{\bf k})$.
Finally, we compare this model, both the formulation as
well as the results, to the problem of interacting electrons in a
diffusive conductor.

\end{abstract}


\maketitle

\section{Introduction}

A particle prepared in a non-equilibrium state 
and interacting with an environment 
usually relaxes to equilibrium. The 
decay of the off-diagonal elements of its density matrix
is denoted as dephasing. More generally,
any state of the particle  described by mixture
with density matrix $\hat\rho \not=\hat\rho^2$ can be interpreted
 as manifestation of dephasing. This applies for reduced
 density matrices, obtained after tracing out the environment,
also in equilibrium and
even in the ground state of the total system. 
We further note that certain correlation and response functions of the particle
decay, a fact which may  be related to dephasing as well. Again this
decay is observable although the expectation values are evaluated in
the ground state of the total system. All these manifestations of
dephasing have the same origin:
the interaction and entanglement of the particle with an
environment, combined with the reduction of the description 
to a subsystem of the total system.
The interaction usually has further consequences, incl.\
relaxation, dissipation, as well as renormalization effects. These
effects, while closely related, must be carefully distinguished
from each other.

In this article we consider several model systems, which in part can be
analyzed exactly, with the idea to illustrate different
manifestations of dephasing and
the distinction to relaxation and renormalization effects.
First we review spin-boson models  \cite{Leggett,Weiss}. Depending
on the spectrum of the bath one finds dephasing, manifest as the
decay of the coherence of an initial non-equilibrium state, 
even at $T=0$. We further show that equilibrium correlation
and response functions decay
on a time scale which coincides with the dephasing time \cite{SMS2}.
On the other hand, some effects of the coupling to the bath
can be interpreted as renormalization effects.
The results are relevant, e.g., in the context of quantum 
manipulations of quantum systems in a dissipative environment  \cite{SMS}.

Next we review Caldeira-Leggett (CL) models  \cite{CL} of a particle coupled
linearly to a bath of oscillators
and arrive at similar conclusions.
As an application we consider
persistent currents in mesoscopic rings with interactions 
 \cite{Buttiker,Florian,Guinea,Herrero,LM}.
We then analyze the problem of a particle which interacts
with a quantum field $\hat V$ as described by the Hamiltonian
\begin{equation}
\hat H = \frac{\hat{{\bf p}}^2}{2m} -e\hat V(t,\hat{\bf r})+ \hat
H_{\rm env} (\hat V, \hat \psi) \; .
\label{Hf}
\end{equation}
The field $\hat V(t,\hat {\bf r})$ fluctuates due to the coupling to
further environment degrees of freedom $\psi$. Its
fluctuations are characterized
by a frequency- and wave-vector-dependent dielectric function
$\epsilon(\omega,{\bf k})$, thus generalizing the effect of the bath
in the CL model. In spite of the differences we find results
similar to those mentioned above, including a finite 
dephasing time at $T=0$.

The results for the dephasing time of the last model coincide with
those derived in Ref.\ \cite{GZ} (GZ) for the problem of interacting
electrons in a diffusive conductor. Indeed, the interaction between
electrons can be accounted for by a fluctuating field $\hat V(t,\hat
{\bf r})$, however, since the electrons are indistinguishable one
has to account for the Pauli principle. An appropriately generalized
formulation of the problem has been presented by GZ \cite{GZ,GZ02}. They
argued that the modifications do not yield qualitative changes for the
dephasing time. This conclusion has been challenged; for a discussion
see, e.g., Refs. \cite{Jan,Florian1}.

\section{Spin-boson model}

The spin-boson model has been  studied extensively before
 \cite{Leggett,Weiss}; the analysis of the following  
section is partially based on work presented in Ref.\ \cite{SMS2}. 
The spin-boson model describes a two-state quantum system coupled
to a bath of oscillators with Hamiltonian 
\begin{eqnarray}
\hat H=-\frac{\Delta E}{2} \, \hat \sigma_z 
+\, \hat X (\cos\theta\,\hat\sigma_z -\sin\theta\,\hat\sigma_x)
+ \sum_k\hbar\,\omega_k\, \hat a^+_k\hat a_k \, .
\end{eqnarray}
Here $\Delta E$ is the bare energy splitting between the levels of the
two-state system. 
The bath operator $\hat X= \sum_k c_k(\hat a_k+\hat a^+_k)$ couples 
`longitudinally' to $\hat\sigma_z$ and `transverse' to $\hat\sigma_x$, 
depending on the angle $\theta$.  
In thermal equilibrium the Fourier transform of the symmetrized
correlation function of this operator,
\begin{equation}
\label{Eq:X-J}
S_X(\omega)
\equiv \left\langle [\hat X(t), \hat X(t') ]_+ \right\rangle_\omega
= 2 \hbar J_s(\omega) \coth \frac{\hbar \, \omega}{2 k_{\rm B}T} \;,
\end{equation}
depends on the bath spectral density 
$J_s(\omega) \equiv (\pi/\hbar)\sum_k c_k^2 \;
\delta(\omega-\omega_k)$.
At low frequencies it typically follows a power-law  
up to a high-frequency cutoff $\omega_{\rm c}$,
\begin{equation}
J_s(\omega) = \frac{\pi}{2}\,\hbar \alpha\, \omega_0^{1-s}\omega^s
\Theta(\omega_{\rm c} - \omega)
\ .
\label{Eq:J_s}
\end{equation}
The spin-boson model has been studied
mostly for baths with Ohmic spectrum ($s=1$).
In general, for dimensional reasons a  frequency scale
$\omega_0$ has been introduced in (\ref{Eq:J_s}), although
$J_s(\omega)$ depends only 
the combination $\alpha\omega_0^{1-s}$. We could choose 
$\omega_0$ equal to the high frequency cutoff $\omega_{\rm c}$ of the
bath~ \cite{Leggett}, however for later discussion   
it is more convenient to distinguish both.

\subsection{Relaxation of a non-equilibrium state}

Two different time scales describe the evolution in the
spin-boson model.  The 
dephasing time $\tau_\varphi$ characterizes the decay of the
off-diagonal elements of the spin's reduced density matrix
$\hat\rho(t)$ in the eigenbasis of $H_{0}$. 
Frequently one encounters an exponential long-time dependence,
$\hat\rho(t)_{12} \sim  e^{-t/\tau_\varphi}$, 
but other decay laws may emerge as well. 
The second, the relaxation time scale $\tau_{\rm relax}$, characterizes how
diagonal entries tend to their thermal equilibrium values,
$\hat\rho_{ii}(t) - \hat\rho^{\rm eq}_{ii} \sim e^{-t/\tau_{\rm relax}}$.
Both times were evaluated in Refs.~ \cite{Leggett,Weiss} with the results
\begin{eqnarray}
\tau_{\rm relax}^{-1}
&=&\frac{1}{\hbar^2} \sin^2\theta \; S_X\left(\omega =
\Delta E/\hbar\right)
\label{Eq:relaxation}
\; , \\ 
\tau_\varphi^{-1}
&=&
\frac{1}{2}\;\tau^{-1}_{\rm relax} + \frac{1}{\hbar^2} \cos^2\theta \, 
S_X(\omega = 0)
\label{Eq:dephasing}
\; .
\end{eqnarray}
For transverse coupling ($\propto \sin \theta$) 
the fluctuating field induces transitions between the
eigenstates of the unperturbed system.   
For longitudinal coupling ($\propto \cos \theta$) it still contributes
to dephasing, since  
it leads to fluctuations of the eigenenergies and, thus, to a
random phase shift.  
This is the origin of the second, ``pure'' dephasing term
$\Gamma^*_\varphi = S_X(\omega=0)/\hbar^2$ in 
Eq.~(\ref{Eq:dephasing}).
For an Ohmic environment at $T\ne 0$  one finds 
$\Gamma^*_\varphi = 2\pi\alpha k_{\rm B}
T/\hbar$. 
At $T=0$, on the other hand, for most spectra the expression
(\ref{Eq:dephasing}) yields a vanishing or divergent
result for  $\Gamma^*$, demonstrating the need
for a more detailed analysis. 

In the limit of purely longitudinal coupling, $\theta = 0$, the
analysis can be done exactly. Assuming a 
factorized initial density matrix one finds $\hat\rho_{12} \sim P_{\omega_{\rm c}}(t)$.
The function $P_{\omega_{\rm c}}(t)$ (known from the
``$P(E)$''-theory \cite{P(E)_Devoret,Nazarov}) 
can be expressed as  $P_{\omega_{\rm c}}(t)=e^{K(t)}$, with
$K(t) = {4\over \pi\hbar} \int_0^{\omega_{\rm c}} d\omega \,
        {J(\omega)\over\omega^2} 
\big[\coth\big({\hbar\omega\over2
        k_B T}\big)(\cos\omega t-1)  -i\sin\omega t\big]$.
For an Ohmic bath ($s=1$), finite temperatures,
and $t>\hbar/k_{\rm B}T$ it reduces to
$ {\rm Re} K(t) \approx 
-\pi\,\alpha\,\frac{2 k_{\rm B}T}{\hbar}\,t$, 
consistent with  Eq.~(\ref{Eq:dephasing}).
On the other hand, for lower temperatures or shorter times,
$1/\omega_{\rm c}<t<\hbar/k_{\rm B}T$,  one finds ${\rm Re}K(t) \approx
-2\,\alpha\,\ln(\omega_{\rm c} t)$, implying a power-law decay 
\begin{equation}
\label{Eq:power_law_dephasing}
\hat\rho_{12}(t)= (\omega_{\rm c} t)^{-2\alpha} e^{- i\Delta E t/\hbar}
\hat\rho_{12}(0)\ .
\end{equation}
Thus even at $T=0$ the off-diagonal elements 
of the density matrix decay in time. It should be noted that all
oscillators up to the cutoff $\omega_{\rm c}$ contribute to
this decay. 

For sub-Ohmic baths ($0<s<1$) with high density of low-frequency 
oscillators exponential dephasing is observed for all temperatures and times: 
$\hat\rho_{12} \propto \exp[-\alpha (\omega_0 t)^{1-s}]$ for
$t<\hbar/k_{\rm B}T$, while $\hat\rho_{12} \propto
\exp[-\alpha \,Tt\,(\omega_0 t)^{1-s}]$ for $t>\hbar/k_{\rm B}T$. 
Thus the dephasing rate is 
$\Gamma_\varphi^* \propto \alpha^{1/(1-s)}\omega_0$ 
for $T<\alpha^{1/(1-s)}\omega_0$ and 
$\Gamma_\varphi^* \propto (\alpha T/\omega_0)^{1/(2-s)}\omega_0$ 
for $T>\alpha^{1/(1-s)}\omega_0$.

In  the super-Ohmic regime ($s>1$) after an initial decay on
time scale $\omega_{\rm c}^{-1}$, 
the exponent ${\rm Re}K(t)$ saturates at a finite value 
${\rm Re}K(\infty) = - \alpha(\omega_{\rm c}/\omega_0)^{s-1}$,  
and the off-diagonal element $\hat\rho_{12}$ stays constant for  
$t < \hbar/k_{\rm B}T$. At longer times, $t>\hbar/k_{\rm B}T$, if
$s<2$ an exponential 
decay is observed, $\hat\rho_{12}(t) \propto
\exp[-\alpha \,Tt\,(\omega_0 t)^{1-s}]$, whereas for $s \ge 2$ there
is almost no additional decay.   

\subsection{Renormalization effects}
Above we assumed factorized initial conditions:
the bath was prepared in the equilibrium state
characterized by temperature $T$, while the spin was 
prepared in an arbitrary initial state. Thus
dephasing is to be expected even at vanishing bath temperature. 
This initial conditions can be achieved in principle by applying
sudden pulses to rotate the spin.  
In a real experiment, however, the preparation pulse 
takes a finite time, $\tau_{\rm p}$, during which the bath 
oscillators partially adjust to the  
changing spin state. This will modify the results for dephasing
 \cite{SMS2}.

For example, a $\pi/2$-pulse, 
which transforms the state $\ket{\uparrow}\to 
\frac{1}{\sqrt{2}} (\ket{\uparrow}+\ket{\downarrow})$, is accomplished by
applying a field
$\hat H_{\rm p}=\hbar\omega_{\rm p}\hat \sigma_x$ for a time $\tau_{\rm
p}=\pi/2\omega_{\rm p}$. 
In this case oscillators 
with high frequencies, $\omega_k \gg \omega_{\rm p}$, 
follow the spin adiabatically, while  
those with low frequency, 
$\omega_k\ll\omega_{\rm p}$, do not change their state. 
Assuming that the oscillators can be split into these two 
groups, we arrive at an initial state where only the low-frequency 
oscillators are factorized from the spin
$\frac{1}{\sqrt{2}}\left(\ket{\uparrow}\otimes\ket{{\rm
g}_{\uparrow}^{\rm h}} + \ket{\downarrow}\otimes\ket{{\rm
g}_{\downarrow}^{\rm h}}\right) \otimes \ket{{\rm g}_{\uparrow}^{\rm
l}}$.
Here the superscripts `h' and `l' refer to  high
and low frequencies  and the 
states $\ket{{\rm g}_{\uparrow/\downarrow}}$ are the 
ground states of the  Hamiltonians 
$\hat H_{\uparrow/\downarrow} \equiv 
\sum_k\hbar\,\omega_k\, \hat a^+_k\hat a_k \pm \hat X$.  

For the off-diagonal element of the density matrix we now obtain 
$\hat\rho_{12} = Z(\omega_{\rm c},\omega_{\rm p})P_{\omega_{\rm p}}(t)$.
Here $Z(\omega_{\rm c},\omega_{\rm p}) \equiv 
|\langle g_{\uparrow}^{\rm h} | g_{\downarrow}^{\rm h}\rangle|$
describes the effect of high-frequency bath oscillators and should be
interpreted as a renormalization, 
while the factor $P_{\omega_{\rm p}}(t) = \bra{{\rm
g}^{\rm l}_{\uparrow}} e^{-i{\cal H}_{\downarrow}t/\hbar} \ket{{\rm
g}^{\rm l}_{\uparrow}}$, 
which reduces to the same form as $P_{\omega_{\rm c}}(t)$ described before 
except that the high-frequency cutoff is reduced to $\omega_{\rm p}$, describes 
dephasing due to low-frequency modes. 

The criterion to distinguish between both is the fact that
renormalization effects are reversible, as illustrated by a
continuation of the Gedanken experiment: After the preparation $\pi/2$-pulse
we allow for a free evolution of the system for some time $t$, 
when the state evolves as $\frac{1}{\sqrt{2}}\Big(e^{i\Delta E
t/2}\ket{\uparrow}\otimes\ket{{\rm 
g}_{\uparrow}^{\rm h}} \otimes \, \ket{{\rm g}_{\uparrow}^{\rm
l}} 
+ \, e^{-i\Delta E t/2} \ket{\downarrow}\otimes\ket{{\rm
g}_{\downarrow}^{\rm h}} \otimes \, e^{-i{\cal
H}_{\downarrow}t/\hbar}\ket{{\rm g}_{\uparrow}^{\rm l}}\Big)$.
Then we apply a $(-\pi/2)$-pulse (also of width $\pi/2\omega_{\rm p}$) and
measure $\hat \sigma_z$. Without  
dissipation the result would be $\langle \hat \sigma_z\rangle =
\cos(\Delta E t)$, with dissipation we obtain  \cite{SMS2}
$\langle\hat \sigma_z\rangle =  
{\rm Re} \left[P_{\omega_{\rm p}}(t) e^{-i\Delta E t}\right]$.
I.e., the amplitude of the coherent oscillations is reduced by the factor 
$|P_{\omega_{\rm p}}(t)|$, associated with slow oscillators.
It describes dephasing, since there is no way to reverse the 
time evolution contained in this factor.
In contrast, the factor $Z(\omega_{\rm c},\omega_{\rm p})$ does not
appear in the final signal. It originates from the overlap of the  
high-frequency oscillator wave 
functions, $\,\ket{{\rm g}^{\rm h}_{\uparrow}}$ and 
$\,\ket{{\rm g}^{\rm h}_{\downarrow}}$. They have been
manipulated adiabatically, and this effect can be reversed.
This effect is properly described by the concept
of renormalization. 

The distinction between both effects can also be demonstrated if 
we discuss the Gedanken experiment using  
renormalized spins, $|\tilde\uparrow\rangle \equiv 
|\uparrow\rangle|g^{\rm h}_\uparrow \rangle$ and
$|\tilde\downarrow\rangle \equiv 
|\downarrow\rangle|g^{\rm h}_\downarrow \rangle$, and reduce the
 high-frequency cutoff of the bath to $\omega_{\rm p}$.

\subsection{Correlation functions}
In the limit $\theta = 0$ we can also calculate exactly the
linear response of $\hat \sigma_x$ to a weak field $H_{1}=-(1/2)
\delta B_x(t) \sigma_x$: 
\begin{equation}
\label{Eq:response function}
\chi(t) = \frac{i}{\hbar}\;\Theta(t) \langle
\left[\hat \sigma_x(t),\hat \sigma_x(0)\right]\rangle \ .
\end{equation}  
Using the equilibrium density matrix
\begin{equation}
\hat \rho^{\rm eq} = \frac{\ket{\uparrow}\bra{\uparrow}\otimes
\hat\rho_{\uparrow} + 
e^{-\beta\Delta E} \ket{\downarrow}\bra{\downarrow}\otimes
\hat\rho_{\downarrow} }{1+e^{-\beta\Delta E}}
 \ ,
\end{equation}
where $\hat\rho_{\uparrow} \propto \exp(-\beta {\cal H}_{\uparrow})$ is the
bath density matrix adjusted to the spin state $\ket{\uparrow}$, and
similar for $\hat\rho_{\downarrow}$, 
we obtain  the imaginary part of the Fourier transform of the
susceptibility, describing dissipation, 
\begin{equation}
\chi''(\omega) = 
\frac{P(\hbar\omega-\Delta E)+ e^{-\beta \Delta E} P(\hbar\omega+\Delta E)
}
{2(1+e^{-\beta \Delta E})}
 - ...(-\omega) \ .
\end{equation}
For an Ohmic bath ($s=1$) at $T=0$ and positive values of $\omega$ we
obtain  \cite{SMS2,P(E)_Devoret,Nazarov} 
\begin{eqnarray}
\chi''(\omega) =
\Theta(\hbar\omega-\Delta E)
\frac{e^{-2\gamma\alpha}(\hbar\omega-\Delta E)^{2\alpha -1}}
{2\Gamma(2\alpha)(\hbar\omega_{\rm c})^{2\alpha}}.
\end{eqnarray}
We observe that the dissipative part
$\chi''$ has a gap $\Delta E$, corresponding to 
the minimum energy needed to flip the spin, 
and a power-law behavior  as $\omega$
approaches the threshold. This behavior of $\chi''(\omega)$
parallels the {\it orthogonality catastrophe}
scenario. It implies that the ground states of the
bath for  different spin states,
$\ket{{\rm g}_{\uparrow}}$ and $\ket{{\rm g}_{\downarrow}}$, are
{\it macroscopically} orthogonal. The form of the response function 
for $s=1$ is also known from the X-ray absorption in metals. 
For sub-ohmic spectra, as $s$ decreases
the $T=0$ shape of $\chi''(\omega)$  gradually evolves
towards a bell shape with width given by the dephasing
rate. At $s=0$ it becomes 
a Lorentzian, which corresponds to the exponential decay of $|P(t)|$ 
in this case.

As $\chi''(\omega)$ characterizes the dissipation
it is interesting to distinguish again the roles of high and 
low frequency oscillators. We use the spectral decomposition 
at $T=0$,
$$
\chi''(\omega) = \pi \sum_\nu |\bra{0} \sigma_x \ket{\nu}|^2
\left[\delta(\omega-E_\nu) - \delta(\omega+E_\nu)\right]\ ,
$$
where $\nu$ denotes the eigenstates of the system. These are 
$\ket{\uparrow} \ket{n_{\uparrow}}$ and 
$\ket{\downarrow} \ket{n_{\downarrow}}$, where 
$\ket{n_{\uparrow/\downarrow}}$  denote the excited
oscillator states 
of the Hamiltonians ${\cal H}_{\uparrow/\downarrow}$.
The ground state is $\ket{\uparrow} \ket{g_{\uparrow}}$, and the 
only excited states contributing to $\chi''(\omega)$ 
are $\ket{\downarrow} \ket{n_{\downarrow}}$ with 
${\cal H}_{\downarrow}\ket{n_{\downarrow}} = 
(\omega - \Delta E)\ket{n_{\downarrow}}$. I.e.\ all 
oscillators with frequencies $\omega_k > \omega - \Delta E$
have to be in the ground state. Thus we find for $\omega_{\rm
p} > \omega - \Delta E$ 
\begin{equation}
\label{Eq:renormalization_of_chi}
\chi_{\omega_{\rm c}}''(\omega) = Z^2(\omega_{\rm c},\omega_{\rm p})
\chi_{\omega_{\rm p}}''(\omega) \, .
\end{equation}
To interpret this result we note that for a model
including a $g-$factor in the coupling to transverse fields,
$H_{1}=-(g/2) \delta B_x(t)  \hat \sigma_x$, the energy absorption is 
proportional to $g^2$. 
Thus, the response function (\ref{Eq:renormalization_of_chi}) of the
spin at frequencies  
$\omega < \omega_{\rm p} + \Delta E$ coincides with the one of a model
with $g=Z(\omega_{\rm c},\omega_{\rm p})$ and cutoff $\omega_{\rm p}$. 
Again, this property of the high-frequency oscillators is naturally
associated with a renormalization phenomenon.

\section{ Particle plus oscillator bath}

In this section we consider an exactly solvable special case of the
Caldeira-Leggett model  \cite{CL}, namely 
a free particle coupled linearly to a bath of oscillators,
\begin{equation}
\hat H=\frac{\hat p^2}{2m}+\sum_k\left[\frac{\hat P_k^2}{2m_k}
+\frac{m_k\omega_k^2}{2}
\left(\hat R_k-\frac{c_k}{m_k\omega_k^2}\hat x\right)^2\right].
\label{HCL}
\end{equation}
The properties of this model strongly depend on the frequency spectrum
of the bath. Here we consider mostly the
Ohmic case with 
\begin{equation}
J(\omega)\equiv
\frac{\pi}{2}\sum_k\frac{c_k^2}{m_k\omega_k}\delta(\omega-\omega_k)
=m \gamma \omega\theta(\omega_{\rm c}-\omega),
\label{spectrum}
\end{equation}
where $\gamma$ is the damping rate in the equation of motion
 derived from (\ref{HCL}) in the classical limit.

\subsection{Equilibrium density matrix at $T=0.$}

For an Ohmic spectrum (\ref{spectrum}) in the ground state of the
total system (particle plus bath)  the reduced density matrix of the
particle  
$\hat \rho(x_1-x_2)=\int dR_k\, \Psi_0(x_1,R_k)\Psi_0^*(x_2,R_k)$,
is found to be
\begin{equation}
\hat \rho(x_1-x_2)=\exp\left[-(x_1-x_2)^2/2L_\varphi^2\right] \; .
\end{equation}
I.e.\ the density matrix decays if $|x_1-x_2|$ exceeds
a certain length $L_\varphi$, which for an Ohmic spectrum is
\begin{equation}
L_\varphi=
\sqrt{\frac{\pi\hbar}{m \gamma \ln(\omega_{\rm c}/\gamma)}}.
\label{Lphi}
\end{equation}

This decay can be observed in equilibrium interference
experiments, e.g., in the
persistent current of a particle in a ring threaded by a magnetic
flux. For a free particle the amplitude of the
current decays with increasing ring radius $R$ as $I \propto
1/mR^2$. If the particle is coupled
to an Ohmic bath, this amplitude decays exponentially  \cite{Guinea,GZ98},
$I \propto \exp(-R^2/L_\varphi^2)$, on the scale given by $L_\varphi$. 

If the bath
spectrum has a gap at low frequencies, then with increasing 
radius the persistent current decreases rapidly, but
beyond some radius it crosses over to the $1/R^2$-dependence  
characteristic for free particles.  Examples are provided
by the models studied in Refs.  \cite{Florian,Guinea,LM}.
In this case,  the effect of the environment can be interpreted as a
renormalization of the particle mass.  
On the other hand, for an Ohmic bath  \cite{Guinea,Herrero,GZ98,Loss} such an 
interpretation is not possible. 
In order to get further insight and to distinguish
between dephasing and renormalization effects one should analyze
the behavior of other physical quantities, such as, e.g., 
fluctuations  \cite{Buttiker} or the real-time decay
of non-equilibrium states.
Below we will provide arguments why we interpret the reduction of the
persistent current as evidence of  
dephasing and justify
denoting the length scale $L_\varphi$ as `dephasing length'.

\subsection{Relaxation of a non-equilibrium state}

Above we illustrated effects of the bath
in the ground state of the total system. To examine the question
whether they are related to dephasing processes,
we consider the relaxation of an excited state. We
start from a factorized initial state
$
\hat\rho_{\rm total}=\hat\rho_{\rm particle}^{(0)}\,\hat\rho_{\rm
bath}(T=0)$, 
where initially the particle is in a
superposition of two  plane waves with opposite momenta:
\begin{equation}
\hat \rho_{\rm particle}^{(0)} = \psi(x_1) \psi^*(x_2),\;\;
\psi(x) =\frac{{\rm e}^{ikx}+{\rm e}^{-ikx}}{\sqrt{2}}.
\end{equation}
The time evolution of the reduced density matrix
of the particle can be expressed by an influence functional  \cite{CL}
$\hat \rho_{\rm reduced}(t,x_{1},x_{2})= \int J(t,x_1,x_2,x_1^{(0)},x_2^{(0)})$
$\times \hat \rho_{\rm particle}^{(0)}(x_1^{(0)},x_2^{(0)}) \, dx_1^{(0)}
dx_2^{(0)}$, which in turn can be written as a path integral
\begin{equation}
J=\int{\cal D}{ x}_1{\cal D}{ x}_2
{\rm e}^{\frac{i}{\hbar}\{S_0[{ x}_1]-S_0[{ x}_2]-
S_{\rm R}[{ x}_1,{ x}_2]+iS_{\rm I}[{ x}_1,{ x}_2]\}}. 
\label{J}
\end{equation}
Here $S_0$ is the action of a free particle, while $S_{\rm R}$ and
$S_{\rm I}$
are associated with  the bath.
For the sake of brevity we do not present explicit
forms of  $J$ and the actions $S_{\rm R/I}.$ We only note
that the path integral (\ref{J}) can be evaluated exactly with the result
 \cite{CL} 
\begin{equation}
\rho_{\rm reduced}(t,x,x)=1+{\rm e}^{-F(t)}\cos[2kx],
\label{decay}
\end{equation}
where
\begin{eqnarray}
F(t)=\frac{2k^2}{\pi m\gamma}\int_0^t dt_1\int_0^t dt_2
\int_0^{\omega_c}d\omega\, \hbar\omega\coth\frac{\hbar\omega}{2T}
\nonumber\\
\times \cos[\omega(t_1-t_2)](1-{\rm e}^{-\gamma t_1})(1-{\rm e}^{-\gamma t_2}).
\label{F(t)}
\end{eqnarray}
The exponent $F(t)$ describes the suppression of
interference, i.e.\ dephasing.
We note that it contains only $\coth\frac{\hbar\omega}{2T},$
rather than the combination $\coth\frac{\hbar\omega}{2T}- 1$
expected from Golden Rule type arguments.
The function $F(t)$ arises as the imaginary part of the action
$S_{\rm I}$ (\ref{J}) evaluated on the saddle-point paths $x_{1,2}$.
It is, thus, affected by $S_{\rm R}$ through 
the damping of these paths, which leads to the factors
$1-\exp[-\gamma t_{1,2}]$ in (\ref{F(t)}).
The role of $S_{\rm R}$ has been discussed recently also 
in Refs \cite{Jan,Florian1}.         

The dephasing time is naturally defined from the condition
$F(\tau_\varphi)=1.$ For long and short times we find
\begin{equation}
F(t)=\frac{4k^2}{\pi m\gamma}\times\left\{
\begin{array}{ll}
(\gamma t)^2\ln\omega_{\rm c}t 
&\mbox{for} \;\;  \gamma t\ll 1, \\
\pi\ln(\omega_{\rm c} t) & \mbox{for} \;\; \gamma t \gg 1.
\end{array}
\right.
\label{asympt}
\end{equation}
Now we distinguish two cases: \\
(1) $k\gg 1/L_\varphi.$ In this case the short-time asymptotic is
sufficient to determine the dephasing time. We find

\begin{equation}
\tau_\varphi\approx \frac{1}{v}
\sqrt{\frac{\pi\hbar}{2m\gamma\ln(\pi\hbar\omega_{\rm c}^2/4m\gamma v^2)}},
\;\; \mbox{with} \;\;
 v=\frac{\hbar k}{m}.
\label{tau1}
\end{equation}
Comparing to Eq.\ (\ref{Lphi}) we observe $L_\varphi\sim
v\tau_\varphi,$ i.e., there exists a simple relation between
the dephasing time associated
with the relaxation of a non-equilibrium state 
and the `dephasing length' $L_\varphi$ found as a ground state
property. 
For later use we also note that $\tau_\varphi$ can be 
obtained directly from $S_{\rm I},$ since for $\gamma t\ll 1$ we
have $F(t)\approx\frac{1}{\hbar} S_{\rm I}(t,vt',-vt').$ 
(In this limit the real part of the action $S_{\rm R}$ (\ref{J})
has no effect on $\tau_\varphi.$ On the other hand, it may be 
important in other contexts, e.g., for the
evaluation of the relaxation rates  \cite{GZ,Florian1}.)\\ 
(2) $k\ll 1/L_\varphi.$ This case is governed by the long-time
asymptotics of $F(t).$ The classical relaxation, which is influenced by 
$S_{\rm R},$ is strong.
The interference pattern decays  as a power-law with 
exponentially long dephasing time
$
\tau_\varphi\approx \omega_{\rm c}^{-1}
\exp\left[\hbar^2m\gamma/(4m^2v^2)\right]
$.
No simple relation between $\tau_\varphi$
and $L_\varphi$ can be established in this limit.

\subsection{Correlation functions}

The dephasing in time can be observed also in the decay of 
equilibrium correlation functions, such as 
$
C(t)=\langle a_1(\hat x(t))a_2(\hat x(0))\rangle
$,
where $a_{1,2}(x)$ are arbitrary functions of coordinates. 
For translationally invariant systems we can express it by Fourier
transforms $
C(t)=\int\frac{dk}{2\pi}\,\tilde a_1(k)\tilde a_2(-k)\,K(t,k)$,
depending on the correlator
\begin{equation}
K(t,k) = \langle{\rm e}^{ik\hat x(t)}{\rm e}^{-ik\hat x(0)}\rangle.
\label{K}
\end{equation}
For a free particle
$K(t,k) = \exp[-\frac{i}{\hbar}\frac{\hbar^2 k^2}{2m}t]$  evolves with
a pure phase factor, while a decay  signals a dephasing process.
In the presence of the bath we find an expression analogous to
that appearing in the $P(E)-$theory  \cite{P(E)_Devoret,Nazarov}
\begin{eqnarray}
K(t,k)=\exp\Big\{-\frac{\hbar k^2\gamma}{\pi m}
\int_0^\infty \frac{d\omega}{\omega (\omega^2+\gamma^2)} \nonumber\\
\times\,\Big({\coth\frac{\hbar\omega}{2T}
(1-\cos\omega t)
+i\sin\omega t}\Big)\Big\} \,.
\end{eqnarray}
At $T=0$ it reduces to
\begin{equation}
|K_0(t,k)|^2\simeq\left\{
\begin{array}{ll}
\exp\big[-\frac{\hbar \gamma k^2}{\pi m}\ln\big(
\frac{1}{\gamma t}\big)t^2\big] & \mbox{for} \, \gamma t\ll 1 \\
\exp\big[-\frac{2\hbar k^2}{\pi m\gamma}(\ln\gamma t + 0.5772..)\big]
 &  \mbox{for} \,\gamma t\gg 1,
\nonumber
\end{array}
\right.
\end{equation}
decaying on the time scale
\begin{equation}
\tau_\varphi\approx\left\{\begin{array}{ll}
(1/v)\sqrt{\pi\hbar/m\gamma\ln[kL_\varphi]} \, &\mbox{for} \;  k\gg
1/L_\varphi \\ 
\gamma^{-1}\exp\{\pi\hbar\gamma/2mv^2\} \,& \mbox{for} \; k\ll 1/L_\varphi.
\end{array}\right.
\end{equation}
Again we find a simple relation $L_\varphi\sim v\tau_\varphi$ for
$k\gg 1/L_\varphi,$ but no simple relation
in the opposite limit.

To illustrate the respective roles of high- and low-frequency oscillators,
it is useful to consider the coupling to only
one oscillator. In this case one finds at $T=0$
\begin{eqnarray}
K(t)=\exp\left\{-\frac{i}{1+\alpha}\frac{\hbar k^2}{2m}t 
-\frac{\alpha\hbar k^2\left(
1-{\rm e}^{-i\Omega t}\right)}{2m\omega(1+\alpha)^{3/2}}\right\} \, ,
\label{Ksingle}
\end{eqnarray}
where $\alpha=c^2/mM\omega^4$ is the coupling strength, and $\omega$ and
$\Omega=\omega\sqrt{1+\alpha}$ are the bare and renormalized frequency
of the oscillator, respectively. 
The function $K(t)$ now displays the phenomenon of beating.  
If the bath contains many oscillators the
beating adds up to an incoherent decay. Clearly
this effect cannot be interpreted in terms
of renormalization of the particle mass. On the other hand,
if the detector is not sensitive to the particle's position
and the frequency of the external field is less than $\Omega,$
then only the motion of the center of mass is important and
one can consider the system ``particle+oscillator'' as a
single particle, analogue to a molecule with the mass $M+m$. This effect
can be regarded as a mass renormalization. Finally, we note that 
the effect of oscillators with very low frequencies can never be
interpreted as a mass renormalization.

\section{Particle in a quantum field}

Next we consider a particle coupled to an
electromagnetic environment as described by the Hamiltonian 
(\ref{Hf}). In the classical limit the fluctuations of the field $V(t,{\bf r})$
are characterized by the dielectric function $\epsilon(\omega,{\bf k})$.
In general the fluctuations are produced by a quantum environment and,
hence, the field should be treated as a quantum field itself. 
Therefore, within the path integral formulation on the Keldysh contour, 
the fluctuating field on the forward part
of the contour $V_1(t,{\bf r})$ is to be distinguished from
the one on the backward part $V_2(t,{\bf r}).$
Both fields are Gaussian distributed with correlators
\begin{eqnarray}
\langle V_i(t,{\bf r})V_j(0,0) \rangle&=&\hbar I(t,{\bf r})
+\frac{i\hbar}{2}[(-1)^{i}R(t,{\bf r})+
\nonumber\\ &&
+\,(-1)^{j}R(-t,-{\bf r})],
\end{eqnarray}
where $ 
I(\omega,{\bf k})  = 
{\rm Im}\left(\frac{-4\pi}{k^2\epsilon(\omega,{\bf k})} \right)
\coth\frac{\hbar\omega}{2T}$ and 
$R(\omega,{\bf k}) = 
\frac{4\pi}{k^2\epsilon(\omega,{\bf k})}$.
We observe that ${\rm Im}(-1/\epsilon(\omega,{\bf k}))$ 
generalizes the spectral density $J(\omega)$ in
CL-like models. In what follows we concentrate on the Drude model of
a normal metal with dielectric function
\begin{equation}
\epsilon(\omega,{\bf k})=\frac{4\pi\sigma}{-i\omega +Dk^2}.
\label{Drude}
\end{equation}
Similar models have been discussed by Weiss \cite{Weiss}, Guinea
 \cite{Guinea,Paco2}, and Cohen \cite{Cohen}.

The evolution of the density matrix can be described again by 
an  influence functional
\begin{eqnarray}
J&=&
\int{\cal D}{\bf r_1}{\cal D}{\bf r_2}\,
{\rm e}^{ \frac{i}{\hbar}\int_0^tdt'\,
\frac{m}{2}(\dot {\bf r}_1^2-\dot {\bf r}_2^2)}
\nonumber\\ &&\times\,
\left\langle
{\rm e}^{ \frac{i}{\hbar}\int_0^tdt'\,
(eV_1(t',{\bf r}_1)-eV_2(t',{\bf r}_2))}
\right\rangle_{V_1,V_2},\nonumber
\end{eqnarray}
which, after averaging over $V_{1,2}$, acquires the form (\ref{J})
where 
(using the notation ${\bf r}'_j={\bf r}_j(t')$)
\begin{eqnarray}
S_{\rm R}=\frac{e^2}{2}\int_0^t \left[R(t'-t'',{\bf r}'_1-{\bf r}''_1)
-R(t'-t'',{\bf r}'_2-{\bf r}''_2)\right.
\nonumber\\
\left. +\,R(t'-t'',{\bf r}'_1-{\bf r}''_2)- R(t'-t'',{\bf r}'_2-{\bf r}''_1)\right] dt' dt'' ,
\nonumber\\
S_{\rm I}=\frac{e^2}{2}\int_0^t  \left[I(t'-t'',{\bf r}'_1-{\bf r}''_1)
+I(t'-t'',{\bf r}'_2-{\bf r}''_2)\right.
\nonumber\\
\left. -\,I(t'-t'',{\bf r}'_1-{\bf r}''_2)- I(t'-t'',{\bf r}'_2-{\bf r}''_1)\right]dt' dt''.
\nonumber
\end{eqnarray}

\subsection{Equilibrium properties.}

The partition function of the system with Hamiltonian (\ref{Hf})
can expressed similar as the corresponding
expressions in the real-time formalism  by a path integral 
$Z\sim \int{\cal D}{\bf r}\,{\rm e}^{-(S_0[{\bf r}]+S_{\rm int}[{\bf r}])},
$
where $S_0[{\bf r}]=\int_0^{\hbar\beta}d\tau\,\frac{m\dot {\bf r}^2}{2}$ and 
\begin{eqnarray}
S_{\rm int}[{\bf
r}]=\frac{e^2}{2}\int_0^{\hbar\beta}d\tau\int_0^{\hbar\beta}ds\,
T\sum_{\omega_\nu} 
\int \frac{d^3{\bf k}}{(2\pi)^3}
\nonumber\\
\times\,\frac{4\pi}
{k^2\epsilon(i|\omega_\nu|,{\bf k})}
{\rm e}^{-i\omega_\nu(\tau-s)+i{\bf k}({\bf r}(\tau)-{\bf r}(s))}.
\end{eqnarray}
As an example we consider again the persistent current of a
particle on a ring with fluctuations characterized by a Drude
dielectric function (\ref{Drude}). 
This problem has been 
addressed in Ref.\ \cite{Herrero}. A decay of the persistent current has
been found for radii exceeding a length scale, $R\gtrsim L_\varphi$, 
\begin{equation}
L_\varphi \sim l (k_Fl)^2.
\label{kFl}
\end{equation}
Here $l$ is the mean free path of electrons in the metal, and $k_F$
the Fermi wave vector.

\subsection{Decay of a non-equilibrium state}

Continuing as in previous sections, we consider the
time evolution of a non-equilibrium state with factorized
initial density matrix. The initial state of the particle is again
assumed to be a sum of two counter-propagating plane waves 
$
\psi_0({\bf r})\propto {\rm e}^{i\bf kr}+{\rm e}^{-i\bf kr}.
$
The visibility of the resulting interference pattern decays in
time as ${\rm e}^{-F(t)}$. The exact solution of this problem is
not possible, but for short times we find
\begin{equation}
F(t) \approx S_{\rm I}[{\bf v}t',-{\bf v}t']/\hbar\approx
t/\tau_\varphi \, ,
\label{Foft}
\end{equation}
where the effective
velocity is ${\bf v}=\hbar {\bf k}/m$, and 
\begin{equation}
\frac{1}{\tau_\varphi}\approx\frac{2 e^2}{v\pi\hbar}
\int_0^\infty\frac{dk}{k}
 \int_0^{kv}d\omega
\,{\rm Im}\left(\frac{-1}
{\epsilon(\omega,{\bf k})}\right)\coth\frac{\hbar\omega}{2T}.
\label{tau3}
\end{equation}
We stress that this form is valid only if the effective
velocity $v$ is high enough, as only in this case the expression
(\ref{Foft}) is applicable in the whole interval
$0<t<\tau_\varphi.$

The Drude formula gives
$
{\rm Im}\left(-1/\epsilon(\omega,{\bf k})\right)= \omega/(4\pi\sigma)
$
for $k<k_{\max}.$ Hence Eq. (\ref{tau3}) reduces for $T=0$ to
\begin{equation}
\frac{1}{\tau_\varphi}=\frac{e^2v k_{\max}^2}{8\pi^2\hbar\sigma}\;, 
\end{equation}
which allows us to define a velocity-independent dephasing length
\begin{equation}
L_\varphi=v\tau_\varphi=\frac{8\pi^2\hbar\sigma}{e^2 k_{\max}^2}.
\end{equation}
If we choose $k_{\max}=1/l,$ where $l$ is the mean free path,
thus accounting for the fact that the Drude expression is not valid at
shorter length scales, we recover (\ref{kFl}).
I.e., as in the previous models, we observe a close relation between
dephasing in real time and the reduction of interference effects in
equilibrium. 

For completeness, we mention that the correlation function
$K(t)=\langle{\rm e}^{ik\hat x(t)}{\rm e}^{-ik\hat x(0)}\rangle$
also decays at the same time scale $\tau_\varphi.$

\section{Interacting electrons}

The model considered in the previous section captures
essential features of the problem of interacting electrons
in disordered conductors. Both the particle
in the example studied above and the electron propagating
in a disordered conductor interact with a quantum field
$\hat V(t,{\bf r})$ (in the latter case produced by all other 
electrons). On the other hand, 
an important and necessary extension is due to the 
property that the interacting electron
is indistinguishable from those producing the fluctuating quantum field,
and the Pauli principle has to be obeyed.
A general path integral formulation for this problem of interacting
electrons, which accounts for the Pauli principle, was
formulated in Ref.\ \cite{GZ} (GZ). Here we do not attempt to repeat this
derivation; we merely proceed with a summary of the main conclusions:

\noindent
The description of GZ yields 
effective actions $S_{\rm R}$ and $S_{\rm I}$ similar to those found
above. More precisely, $S_{\rm I}$ is unchanged,
while the function $R$ in the expression for $S_{\rm R}$ should 
be multiplied by $1-2f(H_0({\bf p}, {\bf r}))$, where
$f(\xi )$ is the Fermi function and $H_0$ the Hamiltonian
of a non-interacting electron. 
This fact has a transparent physical
interpretation. Since at low temperature the factor $1-2f(\xi)$
approaches sign$(\xi -\mu )$ it effectively implies an
energy-dependent 
dissipation: Above the Fermi level ($\xi > \mu$)
the electron can lose energy due 
to the interaction with the bath formed by all other electrons, whereas
the effect of the same bath for  $\xi < \mu $ is to push up 
the holes to the Fermi surface.

The above picture accounts for the difference between the many-body Fermi
system and that of a quantum particle distinguishable from its
environment. However, as GZ stress this difference is unimportant for the
dephasing effect produced by the interaction.  The latter effect 
is due to quantum noise rather than dissipation and it is 
described by the term $S_{\rm I}$, which is not sensitive to the 
Pauli principle. Accordingly,
the expression for $\tau_{\varphi}$ which they derive  \cite{GZ}
coincides with the one found in the previous 
section if one sets $v=v_{\rm F}$. The Pauli principle may influence
the quantum corrections to the classical electron action.
GZ had argued  \cite{GZ} that they only determine the pre-exponent,
and thus are irrelevant for the dephasing time.
On the other hand, von Delft recently conjectured \cite{Jan} that this
effect may be responsible for 
the discrepancy between the conclusion of GZ and others \cite{AA}.
In response, GZ \cite{GZ02} explicitly analyzed the quantum 
corrections, thus confirming their earlier conclusions.

\section{Summary}

We have studied the effect of interactions and coupling to
baths in several model systems. This includes 
spin-boson models, a particle interacting with a Caldeira-Leggett bath,
as well as more general baths with space and time dependent
fluctuation.
In these models, in contrast to the scattering problems discussed by Imry
 \cite{Imry}, we have demonstrated that dephasing effects are observable at 
low temperature in: ($i$) the decay of a 
non-equilibrium initial state, ($ii$) the decay of certain equilibrium 
correlation and response functions, and  ($iii$) equilibrium
properties: e.g., the suppression of persistent currents.

Although details depend on the model, e.g. the spectrum of the
bath, our main conclusions are: (a) In all the models considered
we find a reduction of interference effects due to the coupling to the bath
down to zero temperature. (b) Similarly we find that
dephasing, understood as the decay of non-equilibrium initial states,
persists down to zero temperature. 
(c) Response and correlations functions decay on the same time scales,
even if evaluated in the ground state. 
(d) In some cases the distinction between dephasing and
renormalization effects is ambiguous. An example is the
reduction of equilibrium interference effects. However, we
observe that the 
coupling to the bath which produces the dephasing in nonequilibrium
situations determines in the same combination of parameters the
reduction of interference effects. For this reason we associate these
effects with dephasing as well.
(e) We observe similarities in formulation and results between the
model for a particle in a fluctuating field and the problem of
interacting electrons in disordered metals.

\section*{ACKNOWLEDGMENT}

We thank Y. Makhlin, A. Shnirman, M.~B\"uttiker, F.~Guinea,
C.P.~Herrero, Y.~Imry, D.~Loss, P.~Mohanty,  
M.~Paalanen, J.~von Delft, R.A.~Webb, and U.~Weiss for  
stimulating discussions. 
The work is part of the {\bf CFN} (Center for
Functional Nanostructures) which is supported by the DFG (German Science
Foundation).

\end{document}